# Few Throats to Choke: On the Current Structure of the Internet

H. B. Acharya †[1], Sambuddho Chakravarty †[2], and Devashish Gosain †[2]

[1] Rochester Institute of Information Technology, NY, USA
Email: `acharya@mail.rit.edu`
[2]Indraprastha Institute of Information Technology Delhi (IIITD), New Delhi, India
Email: `{sambuddho, devashishg}@iiitd.ac.in`

†All authors have equal intellectual contributions towards the paper. The author names are therefore arranged in alphabetical order of their last names.

*Abstract*—The original design of the Internet was a resilient, distributed system, that maybe able to route around (and therefore recover from) massive disruption — up to and including nuclear war. However, network routing effects and business decisions cause traffic to often be routed through a relatively small set of Autonomous Systems (ASes). This is not merely an academic issue; it has practical implications — some of these frequently appearing ASes are hosted in censorious nations. Other than censoring their own citizens' network access, such ASes may inadvertently filter traffic for other foreign customer ASes.

In this paper, we examine the extent of routing centralization in the Internet; identify the major players who control the "Internet backbone"; and point out how many of these are, in fact, under the jurisdiction of censorious countries (specifically, Russia, China, and India). Further, we show that China and India are not only the two largest nations by number of Internet users, but that many users in free and democratic countries are affected by collateral damage caused due to censorship by such countries.

## I. INTRODUCTION

Open access to the Internet is an exceptionally powerful political resource. However, several Governments — notably China, Russia, Cuba *etc.* and also some notable democracies such as India, South Africa, and Indonesia, have expressed concern about the open Internet[1]. This concern may be benevolent, *e.g.,* policing child pornography; but there is precedent where state control of communication channels has been abused to silence the opposition. In this context, it is natural to ask how free and open the Internet is — and how robust it is to censorship by these countries.

The Internet, as originally mandated by DARPA, was a resilient communication network that could survive tremendous damage. Internet routing protocols relying on connectionless packet switching help to route around network disruptions. However, in the twenty-plus years that the Internet has been operated by commercial companies, it has evolved and changed. The Internet today consists of more than $50,000$ ASes that peer with one another through complex business relationships (as *customers*, *peers*, or *providers* [2], [3]) so as to forward each others' traffic[2] and as a consequence of business incentives, the Internet now has a hierarchical structure: a few key players intercept a very large fraction of network paths, and potentially, of Internet traffic.

This hierarchically structured Internet might not be a problem, if we are sure that the "key players" are trustworthy. And indeed, the original 2001 study by Subramanian *et al.* [4] that reported that the Internet had a backbone (and presented a five-level hierarchy of the Internet from *dense core* (20 ASes) to *stub customer ASes* (8852 ASes)) indicated that the "core" ASes were major ISPs, from the US, Sweden, and France — countries promoting free online communication. But the fact remains that *the balance of power in the Internet is shifting*. Today, a large fraction of Internet users reside in censorious nations like India, China and Russia [5]; it stands to reason that several key ASes also lie in these nations, and are subject to their censorship policies.

A further concern is the recent report by Houmansadr *et al.* [3] which indicates that, in practice, individuals, companies, and even nations have very limited control over which ASes provide their Internet routes. Even in the case of China, the world's largest nation by number of Internet users (720 million) [5], choosing to re-route around a very small fraction of world ASes would lead to massive and costly disruptions.

1) About 44 ASes in China would have to start functioning as transit ASes (China has only 30 transit ASes, so this is an increase of $150\%$).
2) The effective latency seen by the Chinese user would increase by a factor of 8.

In other words, the hierarchical structure of the Internet *may make it infeasible to perform large scale re-routing*; maintaining Internet access, while avoiding those key ASes of the Internet which act maliciously, is not a viable option for many users.

It is thus natural to ask to what extent the Internet routing policies of key players impact the network access of other ASes, particularly if such key ASes are hosted in *censorious*

---

[1]Kyrgyzstan opposes declaring *access to information* online as a human right. Bangladesh, Congo, and Kenya are opposed to *free speech* online as a human right. Bolivia, Burundi, China, Cuba, Ecuador, India, Indonesia, Qatar, Russian Federation, Saudi Arabia, South Africa, United Arab Emirates, Venezuela, and Vietnam are opposed to both[1].

[2]The existence of such relationships is a constraint on the paths traversed by traffic. For example, if ASes A and B are both providers to AS X, then X will refuse to carry transit traffic from A to B (or B to A).

*nations.* In this regard, our paper looks into the following research questions:

1) **Where are the key ASes today, and how effective are they at capturing Internet traffic?**
   The study by Subramanian *et al.* is fifteen years old; in this time, the Internet has grown from 10,000 ASes to 55,000. How many "heavy-hitter" ASes are there in the current Internet? Are any of these likely censorious? And how much impact can they have?

2) **How much impact do censorious countries have, on network access of other nations?**
   More specifically, are any "heavy-hitter" ASes located in censorious countries also filtering paths which start in other, nearby ASes (legally beyond the censors' jurisdiction), but which pass through the heavy hitters?

To answer the aforementioned questions, we began by mapping the AS-level Internet paths connecting various ASes to popular websites. Our approach, following Gao *et al.* [6], was to use publicly available BGP routes (obtained from various Internet Exchange Points across the globe [7]), and the relationships between the ASes [2], to construct a directed AS-level graph of the Internet, connecting IP prefixes to all ASes of the world.

We observed that very few ASes — 30 (*viz.*, 0.055% of world ASes) — consistently intercepted over 90% of the world paths to the popular websites we chose, whether we took top-10, top-20, ... or top-200 websites (as per Alexa). *Fully one-third of these "heavy-hitter" ASes were found to have their official headquarters in known censorious nations like China, India and Russia*, in stark contrast to the 2001 study where *none* were in censorious countries [4].

A related concern is raised by the observation that Chinese Internet filtering policies inadvertently censor DNS traffic originating outside, but passing through, China [8]. Traffic filtering and monitoring, by the key "heavy-hitter" ASes hosted in a censorious nation, impacts not only "service" traffic from the nation concerned, but "through" traffic from other nations — a clear breach of domain. Our research findings in Section IV reveal that collateral damage is a serious issue for every one of the nine censorious countries we study, and most of all for China (over 92% of all the network paths that traverse Chinese ASes originate outside the network boundaries of China).

As an aside, we also observed that "heavy-hitter" ASes included not only Tier-1 ASes, but a considerable number of Tier-2 ASes, and that, conversely, several Tier-1 ASes were not heavy hitters; this was a surprise to us, given the general folk belief that the "core of the Internet" consists of Tier-1 ASes (discussed in Section V).

## II. BACKGROUND AND RELATED WORK

There are two bodies of work related to the current paper. The first involves the study of censorship and how it is implemented in various countries around the world. The second is the study of Internet mapping, or more precisely, determining the routes taken by Internet traffic. We discuss both of these areas briefly in this section.

### A. Internet Censorship

Government censorship of the Internet was systematically studied by Zittrain and Edelman [9], in their seminal analysis of filtering by the People's Republic of China. Important early studies were then contributed by Deibert *et al.* [10], Wolfgarten [11], and Dornseif [12], who describe not only censorship policy but also mechanism of filtering as well as anti-censorship measures. Work in the area has since focused on either determining exactly which content is blocked in a given country (*i.e.,* policy) or how such blocking is performed (mechanism).

Several prior efforts emphasize on policy. For *e.g.*, authors have explored the censorship in single countries such as China [13], Iran [14], Pakistan [15], *etc.*; Verkamp and Gupta [16] extend this with a survey of censorship across eleven countries. Several projects provide tools to determine censorship policy: ConceptDoppler [17], HerdictWeb [18], CensMon [19], and Encore [20].

Other studies, focusing on mechanism, show a steady increase in the sophistication of both censorship and anti-censorship, from the early work of Clayton [21] *et al.* (TCP reset) and Park and Crandall [22] (HTML response filtering) to the complex arsenal used by China to block Tor, reported by Winter and Lindskog [23]. Our work, in particular, is strongly influenced by two papers in this group: anonymous [8], which raised concerns that *collateral damage* can be caused by the Internet filtering in a nation, and Houmansadr *et al.* [3], who describe the costs of trying to avoid a particular AS.

### B. Internet Mapping

Our work draws heavily on the construction of a map of routes in the Internet. The early work in this area, such as by Govindan and Tangmunrunkit [24], Chang *et al.* [25], and Shavitt and Shir [26], relies on discovering router-level maps using the tool Traceroute, and then uses heuristics to deduce ASes and their connections. However, we make use of the algorithm by Gao [2], which directly extrapolates AS-level paths using public BGP routing data collected by Routeviews [7].

It is reasonable to ask why we did not employ the Traceroute-based approach, *i.e.* mapping the Internet by sending traceroute probes from various vantage points to IPs in different ASes, when it has been employed by the CAIDA Ark Project [27] and iPlane [28]. Our reason is that, while our approach does have limitations (we are limited by the accuracy and completeness of publicly available routing tables), an approach based on Traceroute is even more limited — by the network locations and availability of the (volunteer) probing nodes; and by the accuracy of IP-to-ASN mapping. We, therefore, chose the Gao algorithm, as being more accurate in computing AS-level paths between any two randomly chosen ASes.

More recently, Claffy *et al.* [29] and Luckie *et al.* [30] have demonstrated improved methods of Internet mapping, which are very accurate in deducing AS relationships (provider-customer, peer-peer). We have therefore taken the relationships they compute, and used this information in finding routes in the Internet with Gao's algorithm.

### III. APPROACH AND METHODS

Our primary question, in this paper, is whether a small set of Autonomous Systems actually route all or nearly all of the traffic in the Internet—and if so, to identify these ASes. A high-level overview of our approach is as follows.

1) Collect BGP-level routes in the Internet, to a large set of important targets (such as Google, Facebook, Amazon *etc.*) and construct an AS-level map of the routes connecting all ASes to these destinations.
2) Identify the "heavy-hitter" ASes on the map — those which appear on a large fraction (nearly all) of the traces.
3) Repeat the experiment with different sets of target sites, to check that the given heavy hitters are general, and not an artifact of the chosen list of target sites.

It is natural to question why we do not directly map the traffic-heavy paths of the Internet. Unfortunately, direct information about the magnitudes of traffic flows is not publicly available. As an approximation, we map the paths from all ASes to the most popular websites, which account for much of the traffic in the Internet [31]. This approach does have vulnerabilities — it is quite possible that, for example, we choose the Alexa top-100 websites for our study, and the map we construct is completely different than for the top-200 or some other equally valid set. In order to guard against such a possibility, we perform cross-validation by repeating the experiment with multiple target sets.

We now provide the details of our method.

#### A. Mapping the Internet

As discussed in the previous section, there are two principal methods of mapping the Internet. The first method, as used by tools such as CAIDA's Archipelago [27], involves the active measurement of the network using traceroute *etc.* Probes are sent along various paths, and the hop-by-hop path is computed, then abstracted to AS-level resolution. The second method is to collect publicly-available routing information, from the BGP announcements of ASes, and to collate these routes and extrapolate maps of the Internet.

In this paper, we have adopted the second method. We build an AS-level Internet map, using the paths connecting popular websites and the various ASes of the Internet, using the approach described by Gao *et. al.* [2].

The approach estimates paths from a given IP or IP-prefix to every AS in the Internet. The inputs to the algorithm are existing BGP RIBs; we use the BGP routing tables collected by the RouteViews project [7] from Internet Exchange Points (IXes), where several ASes peer and advertise their available routes.

Paths directly obtained from RIBs are termed *sure paths*. ASes on sure paths are called *Base ASes*. For example, in the (hypothetical) path $2869 - 3586 - 49561 - 58556 - 9829 - 192.0.2.0/24$, each number represents an AS. The path originates at AS2869 and terminates at AS9829, the home AS of the advertised prefix 192.0.2.0/24. Note that the suffixes of sure paths are themselves also sure paths.

In addition to sure paths, the algorithm computes new ones. This is done by extending sure paths to other ASes to which there are no explicitly-known paths (from the prefix concerned). The extended path must be loop-free, and must satisfy the *Valley-Free Property* [2]. The process is as follows.

Our original map uses the top-100 most popular websites (as reported by Alexa) as the target WWW destinations; we then perform cross-validation, to check that our results are not an artifact of these sites (as discussed in detail later in this section).

- For each prefix, all sure paths (containing all the base ASes) are selected. (These are simply the RIB entries corresponding to the input prefix).
  Next, these sure paths are to be inspected for possible extension to new ASes, provided they they satisfy the Valley Free property and have no loops.
- The algorithm searches for ASes that share valid business relations with the current end ASes of paths. (Rather than attempt to infer relationships, we directly used the relationships presented by CAIDA [32].)
- One edge is chosen. It is simply assumed that this edge extends the given sure path by one hop.
  Note that we are trying to find a path from an AS to the target prefix, and that extensions of several sure paths might connect the chosen AS to the prefix. Hence there is a need for tie breaking.
  - The algorithm sorts the possible paths, and selects the *shortest* path to the prefix.
  - In case of a tie, the path with minimum *uncertainty* (length of the inferred path extensions) is chosen.
  - If there is still a tie, the path with the higher *frequency index* (the number of times a sure path actually appears in the RIBs) is selected.
- The frequency with which the chosen edge appears in the RIBs, the uncertainty of the extended path, and the new path length, are updated.

#### B. Identifying ASes of interest

To select ASes of interest from our map of Internet paths, we take a greedy approach. Ranking the ASes by *path frequency* (*i.e.,* how frequently an AS appears on the paths in the graph), we keep selecting the most-frequent ASes until we achieve a desired level of coverage. We choose 90% coverage as our target, *i.e.,* we select enough ASes to give us a cover of at least 90% of the paths in the graph.

It may be questioned here why we do not follow the standard approach of CAIDA [33], where the "importance" of an AS is determined by its customer-cone size (the total number of its customers, customers of customers, *etc.*). In Section V, we show that in fact customer cone size is poorly correlated to path frequency — the actual metric of our interest — and explain why this is so.

#### C. Validation

The most important question regarding our study, is how general its results are. If for example, we find that a small set

341

of "key" ASes dominate routing in the Internet, can we trust this claim, or is it only true for routes to our sample of target sites (Alexa top-100)?

To address this concern, we repeated our experiment for various target sets (Alexa top-10, top-20, top-30 ... top-200 sites) to see if our results remained stable. Finally, we performed direct cross-validation by computing "heavy-hitter" ASes from paths to one set of sites (Alexa top-100) and checking whether they cover over 90% of paths to a different, disjoint set (Alexa ranks 101 to 225).

In this context, we should also consider why we did not simply use our algorithm to plot paths from every AS to every other AS in the world. The reason is that over 85% of the Internet consists of eyeball ASes [3], which primarily consume content from a small number of providers; the overwhelming majority of computed paths in such an all-to-all map would see almost no traffic. Our map of paths from all ASes to important destinations, in contrast, gives a reasonable picture of the actual paths taken by traffic.

## IV. EXPERIMENTAL RESULTS

In this section, we present our experimental results. First, we consider the map constructed with paths to our original sample, the Alexa top-100 test sites. We then check whether our results remain unchanged as we vary the set of target sites in our test.

### A. Test 1 : Alexa top-100

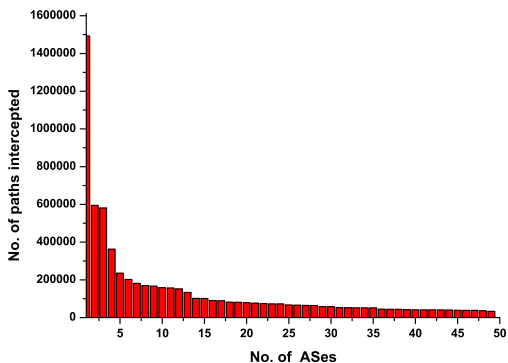

Fig. 1. Paths to Alexa top-100 sites captured by ASes

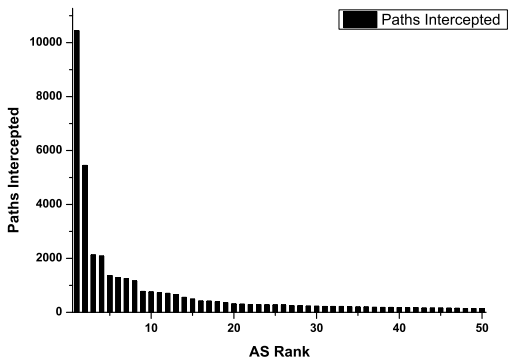

Fig. 2. Paths to one example target site (facebook.com) captured by ASes

The most important result we observe is that the frequency with which "heavy-hitter" ASes appear on paths is remarkably top-heavy, not only for our entire sample of test sites (shown in Figure 1) as an aggregate, but even for the individual sites tested (shown in Figure 2).

The highest-ranked AS, AS3356 (Level 3 Communications), intercepts $1,492,079$ paths ($\approx 33\%$ of total paths).[3] The next highest, AS174 (Cogent Communications), intercepts $536,752$ more paths (not counting overlaps, *i.e.,* paths intercepted by both). Together, AS3356 and AS174 intercept $2,028,831 (= 1,492,079+536,752)$ unique IP-prefix-to-AS paths, *i.e.,* about 45% of all the paths. Proceeding similarly, we see that the top 30 ASes by path frequency together intercept 92.4% of all paths. The complete list is presented in Table III. It represents the ASNs, the country where it is headquartered and their ranks with respect to path frequency ($P_{freq}$) and customer cone size ($C_{size}$). As is clearly visible, nearly a third of these key ASes lie in censorious countries. (If we include AS 6453, Tata America which, while headquartered in the US, actually belongs to an Indian company — *exactly* one-third of the 30 "key" ASes lies in a censorious country.)

As may be expected, out of the censorious countries, the ones with backbone ASes — Russia (11.09% of world paths) and China (7.39% of world paths), as also India (3.08% of world paths) — cover a substantial fraction of the paths in the Internet[4] (see Table I). This is still much smaller than the U.S. (81.82% of world paths), but overall *censorious nations control* 20.73% *of the paths in the Internet.*

This however portrays a different picture compared to the structure of the Internet that was presented in year 2001 [4]. Approximately 20 ASes that constituted the core of the Internet, were hosted in non-censorious nations (see Table IV). Several of these ASes are non-existent now, mostly due to change in business partnerships. Further, newer ASes have gained prominence due to the large fraction of users that they transport. Our results also reflect this massive growth of the Internet – most of the Internet users of the world now reside in Asia; clearly several "heavy-hitter" ASes are hosted in censorious Asian countries.

### B. Cross-Validation

In order to verify the generality of our results, we repeated our experiment for various target sets (Alexa top-10, top-20, top-30 ... top-200 sites). In each case we found the same ASes cover $\approx 90\%$ of paths. Further, the key ASes computed using the Alexa top-100, also capture over 90% of paths to the websites ranked 101 to 225 (Figures 3 and 4). [We add in passing that we also tested how well our "key" ASes covered paths to the 50 most popular non-domestic websites in China, Iran, and Pakistan; they covered $> 90\%$ of these paths as well.]

---

[3]Even this figure underestimates the influence of the company, as another of the 30 key ASes - AS 3549, *i.e.,* Global Crossing—belongs to Level 3.

[4]In comparison, other censorious nations have much less impact: Iran covers 0.69%, Saudi Arabia 0.23%, and Venezuela, Egypt and Pakistan less than 0.15% each.

TABLE I
FRACTION OF AS LEVEL PATHS INTERCEPTED BY VARIOUS COUNTRIES

| Country | Fraction of total paths intercepted |
|---|---|
| RU | 11.09% |
| CN | 7.39% |
| IN | 3.08% |
| IR | 0.69% |
| SA | 0.23% |
| VE | 0.16% |
| EG | 0.12% |
| PK | 0.14% |
| BH | 0.04% |

TABLE II
CORE ASES OF THE INTERNET (AS OF 2001 [4]) AND THE COUNTRIES WHERE THEY WERE HOSTED.

| Country | ASNs |
|---|---|
| US | 1755, 209, 3356, 4006, 3967, 2914 2828, 7018, 3561, 1239, 8918 6453, 3549, 174, 701, 1, 2548 |
| FR | 5511 |
| SE | 1833, 4200 |

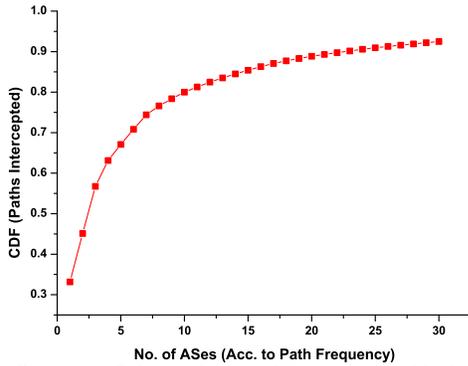

Fig. 3. Cum. freq.: Paths to Alexa top-100 sites captured by key ASes

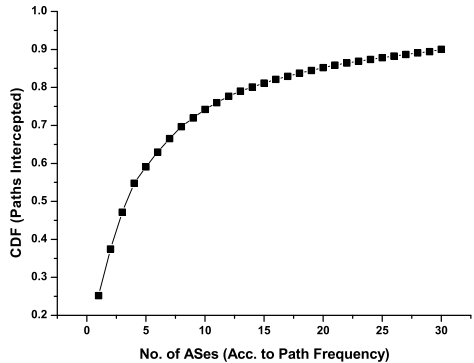

Fig. 4. Cum. freq.: Paths to Alexa sites 101 - 225 captured by the same ASes

TABLE III
THE 30 "KEY" ASES, WHICH INTERCEPT MORE THAN 90% OF PATHS. *ASes headquartered in censorious nations highlighted.*

| ASN | Country | Rank ($P_{freq}$) | Rank ($C_{size}$) |
|---|---|---|---|
| 3356 | US | 1 | 1 |
| 174 | US | 2 | 2 |
| 2914 | US | 3 | 5 |
| 1299 | SE | 4 | 4 |
| 3257 | DE | 5 | 3 |
| 6939 | US | 6 | 13 |
| 6461 | US | 7 | 8 |
| 6453 | US | 8 | 52 |
| 7018 | US | 9 | 17 |
| 10310 | US | 10 | 6 |
| 4134* | CN | 11 | 10 |
| 3549 | US | 12 | 79 |
| 4837* | CN | 13 | 85 |
| 209 | US | 14 | 19 |
| 9002 | UA | 15 | 97 |
| 6762* | IT | 16 | 7 |
| 8359* | RU | 17 | 22 |
| 2828 | US | 18 | 30 |
| 20485* | RU | 19 | 21 |
| 16509 | US | 20 | 9 |
| 9498* | IN | 21 | 18 |
| 4323 | US | 22 | 16 |
| 3216* | RU | 23 | 99 |
| 2497 | JP | 24 | 15 |
| 701 | US | 25 | 12 |
| 12956 | ES | 26 | 65 |
| 37100 | MU | 27 | 23 |
| 4826* | AU | 28 | 26 |
| 12389* | RU | 29 | 67 |
| 1335 | US | 30 | 92 |

*C. Collateral Damage*

Collateral damage results when an AS filters sites, and also causes its customers to lose access [8]. If, for *e.g.*, China was to censor the paths routed through our chosen key ASes, not only would Chinese people would lose access to much of the Internet (and certainly to most popular websites); but also customers of Chinese ASes.

In order to estimate the number of customers that are impacted by such censorship, we inspected the paths through and from nine censorious countries. Figure 5 shows the percentage of paths transiting censorious nations that originate at foreign ASes. Similar representative fractions for eight other censorious nations are also presented in Figure 5.

We see that in the case of China, for *e.g.*, filtering traffic through key ASes would impact many customers, over whom Chinese censorship policies should have no control. $306,874$ AS paths, out of a total of $332,742$ paths involving Chinese ASes and leading to popular destinations, *i.e.,* $92.25\%$ — originate at ASes outside China.

We therefore conclude that the impact of censorship in Russia, China, and India is *not limited to their citizens*. This is especially alarming, considering that China (with $721,434,547$ users) and India (with $462,124,989$) are the largest countries by number of Internet users, and home to roughly one-third of the world Internet-using population; Russia ($102,258,256$ users) is the sixth-largest, and extremely well connected [5].

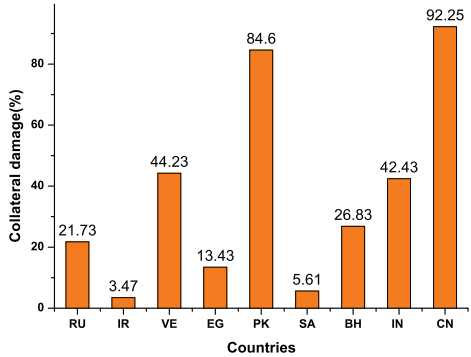

Fig. 5. Ratio of collateral damage (paths filtered that the country *does not have jurisdiction over*) to intentional damage (paths filtered that actually originate in the country), expressed as a percentage.

In Table IV, we show the countries that are potentially impacted by the censorship of China, India and Russia. The list includes most of the major powers: the United States (with $286,942,362$ users), Japan ($115,111,595$), Germany ($71,016,605$), the United Kingdom ($60,273,385$), France ($55,860,330$), *etc.*

TABLE IV
COUNTRIES POTENTIALLY IMPACTED BY COLLATERAL DAMAGE DUE TO FILTERING BY THREE CENSORIOUS NATIONS — CHINA, INDIA AND RUSSIA

| Censorious Country | No. of Countries impacted |
|---|---|
| CN | US, RU, IN, JP, PH, NL, GB, LU, DE |
| IN | US, FR, NL, GB RU, CN, LU, JP, SG, IT, DE |
| RU | US, IN, CN, JP, NL, CH, GB |

## V. DISCUSSION AND FUTURE WORK

From our results in the previous section, it is clear that an overwhelming majority of Internet traffic in our tests (well over 90%) does in fact pass through one or more of a small set of backbone ASes. This would imply that these ASes have the power to set *de facto* censorship policy, and monitor or filter Internet traffic worldwide.

The most important question regarding our work, is how we can claim that this picture is true for Internet traffic *in general*, and not an artifact of our methodology, *i.e.,* that the heavy hitters for flows to Alexa top-100 sites are also heavy hitters for flows to *any* site. We have already discussed our answer to this question in the previous sections, with a description of our cross-validation using different sets of target sites.

In this section, we also describe our finding that the "heavy-hitter" ASes are not necessarily the Tier-1 ASes.

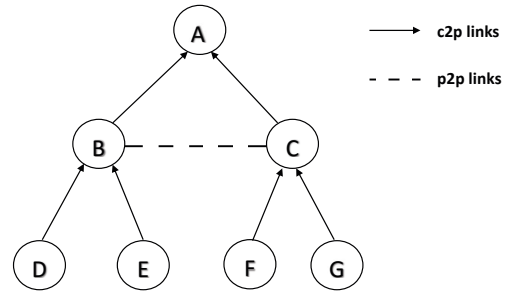

Fig. 6. Schematic AS graph. *A* is "root" of customer cone.

### A. Heavy-Hitters vs Tier-1 ASes

One of the surprising observations of our paper is that the "heavy-hitters" of the Internet not only form a small core, but the size of the core is not much larger than the 20 ASes reported by Subramanian *et al.*, despite the dramatic growth of the Internet in the intervening fifteen years [4].

Another, and perhaps equally surprising, fact is that the "heavy-hitter" ASes we identify are not necessarily Tier-1 ASes (defined as those with only peering relationships, and no providers). For example, our list includes the major Tier-2 ASes Cogent Communications (AS 174) and Hurricane Electric (AS 6939), as well as the ChinaNet backbone (AS 4134 and AS 4837), RosTelecom (AS 12389), Yahoo! (AS 10310) *etc.* which are not only Tier-2 but have Tier-2 providers (Cogent (AS 174) is a provider to RosTelecom, nLayer Communications (AS 4436) to the ChinaNet backbone, and Hurricane Electric (AS 6939) to Yahoo!) We did not, however, observe any Tier-3 ASes. On the other hand, our list does not include five of the sixteen Tier-1 ASes, specifically Deutsche Telekom AG (AS 3320), KPN International (AS 286), Orange (AS 5511), Liberty Global (AS 6830), and Sprint (AS 1239).

We therefore find that the assumption that Tier-1 ASes are the "heavy-hitters" of Internet traffic, is not quite true; there is certainly a strong positive correlation between being Tier-1 and being a "key AS" of the Internet — by which we mean an AS able to intercept most Internet traffic — but it is neither necessary, nor sufficient.

Next, we observed that while many of the ASes on our list were in fact Tier-2, they were very highly ranked by CAIDA [33] in terms of Customer Cone size. This naturally raised the question of whether perhaps a composite feature — Tier-1 *or* large customer cone — would predict if an AS is in fact a key AS *w.r.t.* intercepting Internet traffic. However, there are counter examples for this as well, such as RETN (AS 9002) and SOVAM (AS 3216).

We then experimentally checked whether customer cone size is a good predictor of path frequency. Our results were very surprising: in fact, among our key ASes, the Spearman's Rank Correlation Coefficient between cone size and path frequency is only $\approx 0.2$. We believe the explanation for this result comes from the existence of *non-root paths* in a customer cone, which we explain with the help of Figure 6.

The figure represents a hypothetical AS graph where node *A* is the "root" AS. *A* has the highest customer-cone size in this figure (6 ASes - $D, B, E, F, C, G$) [5]. ASes $B$ and $C$ have customer cones of size 2. Many valid (valley-free) paths — such as $D - B - E$, $D - B - C - F$, $D - B - C - G$, $D - B - A - C - F$, $D - B - A - C - G$, $E - B - A - C - F$ and $E - B - A - C - G$ — do *not* pass through the *root AS*, *i.e.,* the node with the highest customer-cone size.

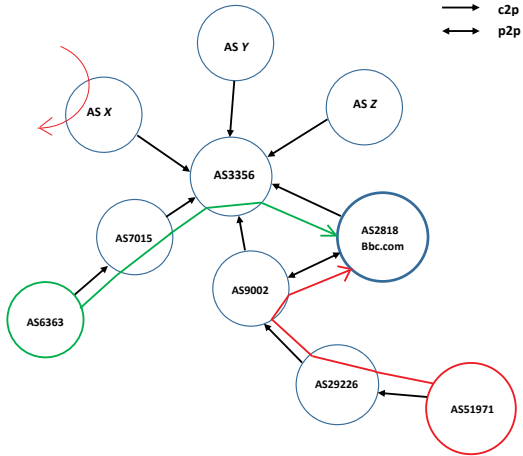

Fig. 7. Valley free paths in the cone of AS3356. Green line: network path that traverses AS3356 to reach AS2818 directly. Red lines: network paths that traverse the one-hop customers of AS3356, but not AS3356 itself.

Our map of the Internet shows that this is indeed a common phenomenon. For example, $34.16\%$ of the paths to top-100 IP prefixes traverse the AS with the largest customer cone, AS3356 (cone size = $24,553$). But nearly as many paths, $33.17\%$, prefer to pass through its 1-hop (immediate) customers. For *e.g.,* as we see in Figure 7, the traffic through AS9002 to AS2818 (www.bbc.co.uk) does not pass through AS3356, though it is the provider to both these ASes. Still more paths pass through $n$-hop customers of root ASes (*i.e.* customers of customers, and so on.) As a result, *customer-cone sizes and AS path frequencies are not well correlated.*

TABLE V
FRACTION OF TRAFFIC PATHS IN A CUSTOMER CONE TRAVERSING LARGE "ROOT" AS, VS FRACTION TRAVERSING 1-HOP CUSTOMERS INSTEAD.

| ASN | % of path not reaching the AS | % of path reaching the AS |
|---|---|---|
| 3356 | 34.16 | 33.17 |
| 174 | 29.05 | 13.13 |
| 2914 | 28.16 | 12.90 |
| 1299 | 36.50 | 8.05 |
| 3257 | 21.00 | 5.23 |
| 6939 | 7.46 | 4.40 |
| 6461 | 5.13 | 4.03 |
| 6453 | 26.00 | 3.76 |
| 7018 | 7.40 | 3.70 |
| 10310 | 0.07 | 3.52 |

[5]The customer cone consists of all the ASes that A can reach via its customers, their customers, *etc.*

We conclude that path frequency is not as strongly correlated with customer cone size as we expected, owing to the considerable fraction of paths that do not transit ASes with large cone sizes (preferring to pass through their customer ASes instead). However, for ASes with smaller customer-cones, we observed fewer such non-root paths (possibly because an AS in a small cone tends to have fewer peers to route through). We may perform a more extensive analysis of such behavior in future work.

### B. Current and Future Work

The primary idea that motivates this work is to map the Internet, and determine which entities (companies and governments) hold the strategic "high ground" of cyberspace. We explore a complementary research direction in another work:

- The largest nation on the Internet by users, China, is highly censorious. India, the second-largest, is rapidly becoming censorious as well. If in future a Great Firewall of India is built along the same lines as the Great Firewall of China, what might it look like, and what mechanisms might it employ? We study this question in our submitted paper [34].

Our results indicate that routing in the Internet is indeed dominated by a few heavy hitters, who therefore enjoy a surprising amount of power. However, several other players in the current Internet economy may also be considered "central" to the Web — the major websites themselves (especially the ones who serve as a platform - most prominently Google and Amazon); root DNS servers; and the major Internet Exchanges (DE-CIX, AMS-IX, LINX, IX.br, DATA-IX and MSK-IX, NL-IX, Equinix, *etc.*) The general question, "who holds the high ground," is thus just as complicated for cyberspace as for the physical world. (The question is very similar to asking: is it the player who controls oil wells who is in a strong strategic position? Or the one with the critical ports on trade routes?) We intend to explore this research direction in detail, in the course of our future work.

### VI. CONCLUDING REMARKS

The organic growth of the Internet has led to a structure that concentrates substantial routing power in a small number of ASes. Our paper experimentally validates this "folk wisdom", and demonstrates that it still holds true even though the Internet has grown and expanded dramatically in the fifteen years since it was first discovered [4]. However, the *main contribution* of our paper is to draw attention to the potential for censorship in this top-heavy structure. A third of the 30 key ASes that form the "heavy-hitter" ASes of the Internet lie in censorious countries (unlike what was presented in 2001 [4]), and they cover over $20\%$ of the Internet paths in our tests. Moreover, from direct examination we see that censorious countries filter (and possibly also monitor) a *substantial* fraction of traffic from other countries.

Further, we also discover that the "key" ASes of the Internet, who carry the overwhelming majority of traffic, are not identical to the Tier-1 ASes (as might be expected from the colloquial use of "Tier-1").

We conclude that while it is certainly understandable that the more powerful routing companies successfully increase their influence over time, perhaps such centralization is effectively making the Internet more fragile as it leads to a small number of "throats to choke". We will pursue this direction further in our future work.

## REFERENCES

[1] Article19, "Significant resolution reaffirming human rights online adopted." [Online]. Available: https://www.article19.org/resources.php/resource/38429
[2] L. Gao, "On inferring autonomous system relationships in the internet," *IEEE/ACM Trans. Netw.*, vol. 9, no. 6, pp. 733–745.
[3] A. Houmansadr, E. L. Wong, and V. Shmatikov, "No direction home: The true cost of routing around decoys." in *NDSS*, 2014.
[4] L. Subramanian, S. Agarwal, J. Rexford, and R. H. Katz, "Characterizing the internet hierarchy from multiple vantage points," in *INFOCOM 2002. Twenty-First Annual Joint Conference of the IEEE Computer and Communications Societies. Proceedings. IEEE*, vol. 2. IEEE, 2002, pp. 618–627.
[5] "Internet live statistics." [Online]. Available: http://www.internetlivestats.com/
[6] J. Qiu and L. Gao, "As path inference by exploiting known as paths," in *Global Telecommunications Conference, 2006. GLOBECOM'06. IEEE*. IEEE, 2006, pp. 1–5.
[7] D. Meyer, "The university of oregon routeviews project." [Online]. Available: www.routeviews.org
[8] Anonymous, "The collateral damage of internet censorship by dns injection," *SIGCOMM Comput. Commun. Rev.*, vol. 42, no. 3, pp. 21–27.
[9] J. Zittrain and B. Edelman, "Internet filtering in china," *IEEE Internet Computing*, vol. 7, no. 2, pp. 70–77, 2003.
[10] R. Deibert, J. Palfrey, R. Rohozinski, J. Zittrain, and J. G. Stein, *Access denied: The practice and policy of global internet filtering*. Mit Press, 2008.
[11] S. Wolfgarten, "Investigating large-scale internet content filtering," *Online: http://www. devtarget. org/downloads/dcu-mssf-2005-wolfgarten-filtering. pdf [2 5.07. 2008]*, 2006.
[12] M. Dornseif, "Government mandated blocking of foreign web content," *arXiv preprint cs/0404005*, 2004.
[13] J. Wright, "Regional variation in chinese internet filtering," *Information, Communication & Society*, vol. 17, no. 1, pp. 121–141, 2014.
[14] S. Aryan, H. Aryan, and J. A. Halderman, "Internet censorship in iran: A first look." in *FOCI*, 2013.
[15] Z. Nabi, "The anatomy of web censorship in pakistan." in *FOCI*, 2013.
[16] J.-P. Verkamp and M. Gupta, "Inferring mechanics of web censorship around the world." in *FOCI*, 2012.
[17] J. R. Crandall, D. Zinn, M. Byrd, E. T. Barr, and R. East, "Conceptdoppler: a weather tracker for internet censorship." in *ACM Conference on Computer and Communications Security*, 2007, pp. 352–365.
[18] J. Zittrain, R. Budish, and R. Faris, "Herdict: Help spot web blockages," 2014.
[19] A. Sfakianakis, E. Athanasopoulos, and S. Ioannidis, "Censmon: A web censorship monitor," in *USENIX Workshop on Free and Open Communication on the Internet (FOCI)*, 2011.
[20] S. Burnett and N. Feamster, "Encore: Lightweight measurement of web censorship with cross-origin requests," *ACM SIGCOMM Computer Communication Review*, vol. 45, no. 4, pp. 653–667, 2015.
[21] R. Clayton, S. J. Murdoch, and R. N. Watson, "Ignoring the great firewall of china," in *International Workshop on Privacy Enhancing Technologies*. Springer, 2006, pp. 20–35.
[22] J. C. Park and J. R. Crandall, "Empirical study of a national-scale distributed intrusion detection system: Backbone-level filtering of html responses in china," in *Distributed Computing Systems (ICDCS), 2010 IEEE 30th International Conference on*, 2010, pp. 315–326.
[23] P. Winter and S. Lindskog, "How the great firewall of china is blocking tor." in *FOCI*, 2012.
[24] R. Govindan and H. Tangmunarunkit, "Heuristics for internet map discovery," in *INFOCOM 2000. Nineteenth Annual Joint Conference of the IEEE Computer and Communications Societies. Proceedings. IEEE*, vol. 3. IEEE, 2000, pp. 1371–1380.
[25] H. Chang, S. Jamin, and W. Willinger, "Inferring as-level internet topology from router-level path traces," in *ITCom 2001: International Symposium on the Convergence of IT and Communications*. International Society for Optics and Photonics, 2001, pp. 196–207.
[26] Y. Shavitt and E. Shir, "Dimes: Let the internet measure itself," *ACM SIGCOMM Computer Communication Review*, vol. 35, no. 5, pp. 71–74, 2005.
[27] K. C. Claffy, "The caida archipelago project." [Online]. Available: http://www.caida.org/ projects/ark/
[28] H. V. Madhyastha, T. Isdal, M. Piatek, C. Dixon, T. Anderson, A. Krishnamurthy, and A. Venkataramani, "iplane: An information plane for distributed services," in *Proceedings of the 7th symposium on Operating systems design and implementation*. USENIX Association, 2006, pp. 367–380.
[29] K. Claffy, Y. Hyun, K. Keys, M. Fomenkov, and D. Krioukov, "Internet mapping: from art to science," in *Conference For Homeland Security, 2009. CATCH'09. Cybersecurity Applications & Technology*. IEEE, 2009, pp. 205–211.
[30] M. Luckie, B. Huffaker, A. Dhamdhere, V. Giotsas, and K. C. Claffy, "As relationships, customer cones, and validation," in *Proceedings of the 2013 Conference on Internet Measurement Conference*, pp. 243–256.
[31] "100 websites that rule the internet," https://www.vodien.com/da-100-websites-rule-internet.php.
[32] "AS Relationships," http://www.caida.org/data/as-relationships/.
[33] "What is cone size?" [Online]. Available: http://www.caida.org/data/as-relationships
[34] D. Gosain, A. Agrawal, S. Sekhawat, H. B. Acharya, and S. Chakravarty, "Mending wall: On the implementation of censorship in india," in *Proceedings of the 13th EAI International Conference on Security and Privacy in Communication Networks (SECURECOMM)*, 2017.